\documentclass[]{PHYEAUTH}
\usepackage{epsfig}

\begin{document}
\begin{frontmatter}

\title{
Ground state phase diagram of 2D electrons in high magnetic field
}

\author{Naokazu Shibata\thanksref{thank}}
and
\author{Daijiro Yoshioka}

\address{Department of Basic Science,
University of Tokyo, Komaba 3-8-1, Tokyo 153-8902, Japan}

\thanks[thank]{
Corresponding author. 
E-mail: shibata@phys.c.u-tokyo.ac.jp}

\begin{abstract}
The ground state phase diagram of two-dimensional 
electrons in high magnetic field is
studied by the density matrix renormalization group 
(DMRG) method. The low energy excitations and pair 
correlation functions in Landau levels
of $N=0,1,2$ are calculated for wide range of
fillings. The obtained results for systems with up to 
25 electrons confirm the existence of various 
electronic states in quantum Hall systems. 
The ground state phase diagram for $N=0,1,2$
consisting of incompressible liquids, compressible liquids,
charge density waves called stripe, bubble and Wigner 
crystal is determined.
\end{abstract}

\begin{keyword}
Phase diagram, Stripe, Bubble, Wigner crystal, Laughlin state, DMRG
\PACS 73.43.Cd \sep 71.10.Pm \sep 73.20.Qt \sep 73.40.Kp 
\end{keyword}
\end{frontmatter} 

\section{Introduction}
In ideal two-dimensional systems the kinetic energy of
electrons is completely quenched by 
perpendicular magnetic field and the macroscopic
degeneracy in the partially filled Landau level is lifted 
only by Coulomb interaction.
The Coulomb interaction projected onto the lowest Landau 
level has strong short range repulsion 
and the Laughlin state is realized at $\nu=1/(2n+1)$\cite{Lagh}.
In the limit of low filling $\nu\rightarrow 0$, however,
the electrons are equivalent to classical point particles 
and the ground state is Wigner crystal.
There exists competition between the quantum liquid
and Wigner crystal, 
and the determination of the phase diagram with respect to
the filling factor $\nu$ has been an interesting subject.
On the other hand, in higher Landau levels, the short range 
repulsion is reduced because one-particle eigenstates of 
the electrons extend over space. 
The Laughlin state is no longer realized in high Landau levels.
Instead, CDW states called stripe and bubble are predicted
in the Hartree-Fock theory\cite{Kou1}.
Indeed, anisotropic resistivity consistent with the formation 
of the stripe state has been observed experimentally\cite{Lill}. 
Thus there is another interesting evolution of the ground state 
when we increase the Landau level index $N$.

In the present study, we investigate the ground state 
of 2D electrons for wide range of filling
by using the density matrix renormalization group (DMRG) method,
and determine the phase diagram for Landau levels of $N=0,1,2$.
Part of the results has already been published\cite{Shib1,Shib2}.
Here we further investigate the stability of the stripe state in the 
lowest Landau level around $\nu=3/8$\cite{Shib2},
whose period of the stripes is much shorter than that
obtained in the composite fermion theory\cite{CFBu},
and show that a transition to a liquid state takes place
when the width of two-dimensional
system becomes several times larger than the magnetic length.

\section{Model and method}
We use the Hamiltonian of the ideal 
two-dimensional system in a perpendicular magnetic field, 
which contains only Coulomb interaction,
\begin{equation}
H=\sum_{i<j} \sum_{\bf q} e^{-q^2/2} \left[ L_{N}(q^2/2) \right] ^2 V(q) 
e^{i{\bf q} \cdot ({\bf R}_i-{\bf R}_j)},
\label{Coulomb}
\end{equation}
where the Coulomb interaction is projected onto the Landau level 
of index $N$, and ${\bf R}_i$ is the guiding center 
coordinate of the $i$th electron, which satisfies the 
commutation relation,
$[{R}_{j}^x,{R}_{k}^y]=i\ell^2\delta_{jk}$.
$L_{N}(x)$ are the Laguerre polynomials
and the magnetic length $\ell$ is set to be 1.
We assume completely spin polarized ground state
and neglect the electrons in fully occupied Landau levels.

In order to determine the ground state phase diagram, 
we need to systematically calculate the ground state 
at various fillings $\nu$ without any artificial biases.
For this purpose we employ the DMRG method\cite{DMRG}, which
enables us to obtain essentially the exact ground state 
for large size of systems extending the limitation of the
exact diagonalization method.
In the DMRG method we iteratively expand the size of system
by adding new local orbitals with
restricting the number of basis states 
using the density matrix calculated from the ground state 
wave function.
The truncation error in the DMRG calculation is estimated from the 
eigenvalues of the density matrix and accuracy of the results
is systematically improved by increasing the number of basis 
states retained in the system.

In the present study we calculate the ground state
for various size of systems with up to 25 electrons
in the unit cell of $L_x\times L_y$ with periodic boundary 
conditions in both $x$ and $y$ directions. 
We choose the aspect ratio $L_x/L_y$ by searching an
energy minimum with respect to $L_x/L_y$ 
to avoid artificial determination of CDW structure.
The truncation error in the DMRG calculation is
typically $10^{-4}$ for 25 electrons with 180 states in each block.
The existing results of exact diagonalizations are 
completely reproduced within the truncation error.
Since the present Hamiltonian has the particle-hole symmetry, 
we only consider the case of $\nu_N \le 1/2$,
where $\nu_N$ is the filling factor of partially filled 
Landau level of index $N$.

We identify the ground state 
by analyzing the low energy excitations and
the pair correlation functions defined by
\begin{equation}
g({\bf r}) \equiv \frac{L_x L_y}{N_e(N_e-1)}\langle 
\Psi | \sum_{i\neq j} \delta({\bf r}+{\bf R}_i-{\bf R}_j)|\Psi
\rangle,
\end{equation}
where $|\Psi\rangle$ is the ground state.

\begin{figure}[t]
\begin{center}\leavevmode
\epsfxsize=70mm \epsffile{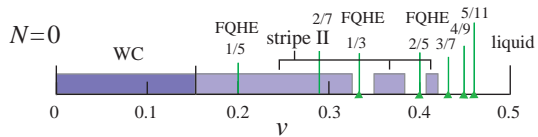}
\caption{The ground state phase diagram of the lowest
Landau level obtained by the DMRG calculation\cite{Shib2}.}\label{n0}
\end{center}\end{figure}


\section{Results}
\begin{figure}[t]
\begin{center}\leavevmode
\epsfxsize=70mm \epsffile{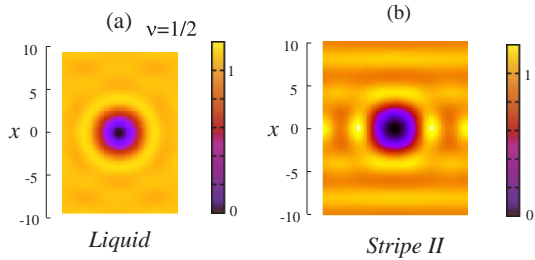}
\caption{Pair correlation functions $g({\bf r})$ of 
(a) compressible liquid state at $\nu=1/2$, 
(b) type-II stripe state at $\nu=0.37$
in the lowest Landau level.}
\label{n0gr}
\end{center}\end{figure}

In the lowest Landau level, the pair correlation functions and the 
size dependence of the excitation gap
show the existence of fractional
quantum Hall states at various $\nu$\cite{Jain,Halp}
as shown in Fig.~\ref{n0}. 
We have calculated low energy excitations of various systems with up to 
25 electrons and confirmed that the excitation gap
at $\nu=n/(2n+1)$ decreases with increasing $n$.
At $\nu=1/2$ the size dependence of the excitation gap
is different from that observed at $\nu=n/(2n+1)$,
and the gap seems to vanish in the limit of large system\cite{Shib2}.
The pair correlation function at $\nu=1/2$
confirms liquid ground state as shown in Fig.~\ref{n0gr} (a).
The effective mass $m^*$ of composite fermions 
estimated from the size dependence of the excitation gap
is about $5$ in units of $\varepsilon \ell/e^2$, which is
almost consistent with the previous study
of the exact diagonalizations on sphere geometry\cite{Onod}.

\begin{figure}[t]
\begin{center}\leavevmode
\epsfxsize=55mm \epsffile{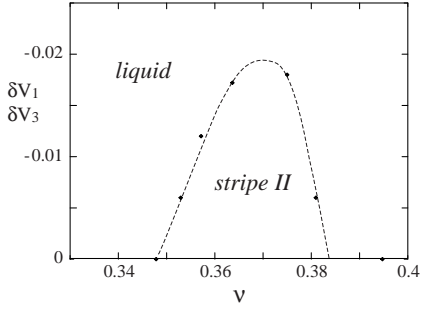}
\caption{Phase diagram of type II stripe state
around $\nu=3/8$ in the lowest Landau level.
$\delta V_1= \delta V_3=0$ corresponds to the ideal 2D system.
$\delta V_1$ and $\delta V_3$ are chosen to be the same.
The broken line is a guide for the eyes.
}
\label{st_phase}
\end{center}\end{figure}

Below $\nu\sim 0.42$, we find CDW states between the
incompressible states at $\nu=n/(2n+1)$. 
As shown in Fig.~\ref{n0gr} (b),
the pair correlation function has stripe
structure\cite{period} but it shows clear oscillations from the origin
along the stripe.
This structure is different from the
stripes in higher Landau levels near half filling,
and we refer this new state as type II stripe state
(stripe II) \cite{Shib2}.

In order to investigate the stability of this
stripe state, we slightly decrease the 
short range components of the effective interaction
represented by the Haldane's pseudo potentials,
$V_1$ and $V_3$ \cite{pseudo}.
The obtained ground state around $\nu=3/8$ is summarized
in Fig.~\ref{st_phase}.
With decreasing $V_1$ and $V_3$,
we find a transition to a liquid state.
Such decrease in the short range pseudo potentials
is realized in real systems, which have finite width
perpendicular to the two-dimensional plane.
Since a few percent decrease $(\delta V_3 \sim -0.02)$ corresponds 
to a width of several times the magnetic length \cite{LDA},
the stripes in the lowest Landau level are not realized 
in wide quantum wells and under high magnetic fields.
This stripe state may be related to the 
reentrant insulating phase above $\nu = 1/3$ recently 
observed in a narrow quantum well\cite{narrow}.

With further decreasing $\nu$,
the CDW structure is enhanced, and 
the first order transition to Wigner crystal
takes place at $\nu\sim 1/7$\cite{Shib2,Lam}.
At $\nu=1/5$ the short range pair correlation function
is similar to that of the CDW state around  $\nu=1/5$ but
the ground state continuously connects to 
the Laughlin state, which is an exact ground state in the limit of large 
$V_1$ and $V_3$\cite{Shib2}.

\begin{figure}[t]
\begin{center}\leavevmode
\epsfxsize=70mm \epsffile{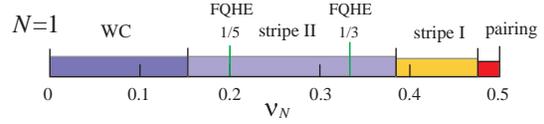}
\caption{The ground state phase diagram of the second-lowest
Landau level obtained by the DMRG calculation\cite{Shib2}. }\label{n1}
\end{center}\end{figure}

\begin{figure}[t]
\begin{center}\leavevmode
\epsfxsize=75mm \epsffile{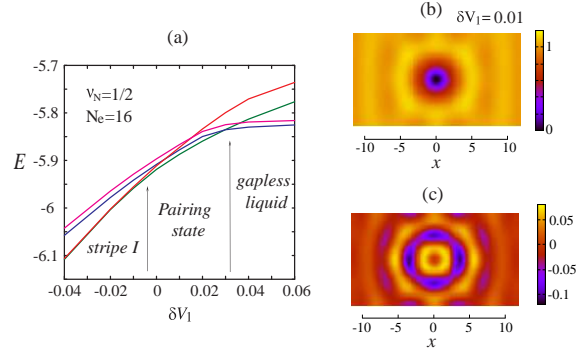}
\caption{(a) Low energy spectrum at $\nu_N=1/2$
in the second lowest Landau level.
(b) Pair correlation function at  $\delta V_1=0.01$.
(c) Difference in $g({\bf r})$ between pairing state at $\delta V_1=0.01$
and gapless liquid state at $\delta V_1=0.06$,
$g({\bf r})_{\delta V_1=0.01}-g({\bf r})_{\delta V_1=0.06}$.
}\label{pair}
\end{center}\end{figure}

In the second lowest Landau level, various ground states
are obtained as shown in Fig.~\ref{n1}.
At $\nu_N= 1/2$, we have systematically calculated low energy excitations and 
pair correlation functions for various Haldane's pseudo-potentials. 
As shown in Fig.~\ref{pair} (a), the ground state around $\delta V_1=0$ 
is different from neither stripe state in higher Landau levels nor
compressible liquid state in the lowest Landau level.
The pair correlation function is consistent with the pairing 
formation\cite{RezHal,Morf}
as shown is Figs.~\ref{pair} (b) and (c), where
Fig.~\ref{pair} (c) shows the difference in $g({\bf r})$ from gapless 
liquid state at $V_1=0.06$, 
$g({\bf r})_{\delta V_1=0.01}-g({\bf r})_{\delta V_1=0.06}$. 
Between $\nu_N \sim 0.47$ and $0.38$ the stripe ground state is obtained. 
The correlation function is similar to that observed in higher Landau 
levels, which we call type I stripe state, 
but the amplitude of the stripes is 50\% smaller and 
the period is 30\% shorter than those observed in $N=2$ Landau level. 
Below $\nu_N =0.38$, the ground
state correlation functions become similar to the type II stripe state
obtained in the lowest Landau level.
At $\nu_N=1/3$ the ground state correlation function seems to be
quite different from that of the Laughlin state in the lowest Landau
level. However the ground state continuously connects to the Laughlin
state\cite{Shib2}. 
The excitation gap is very small and the pair correlation 
function near the origin is enhanced from that of the Laughlin state. 
This is due to the large reduction of
the short range repulsion $V_1$ in the second lowest Landau level. 
On the other hand, the ground state at $\nu_N=1/5$ is almost 
identical to that in the lowest Landau level, because $V_3$ 
in the second lowest Landau level is slightly larger than
that of the lowest Landau level.
The ground state at low fillings 
is almost the same to that in the lowest Landau level, and the 
first order transition to Wigner crystal is expected at 
$\nu_N \sim 1/7$\cite{Shib2}.
 
\begin{figure}[t]
\begin{center}\leavevmode
\epsfxsize=70mm \epsffile{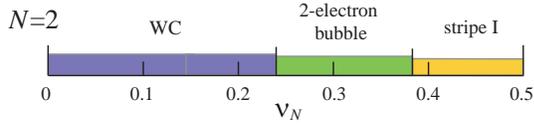}
\caption{The ground state phase diagram of the third-lowest
Landau level obtained by the DMRG calculation\cite{Shib1}. }\label{n2}
\end{center}\end{figure}

\begin{figure}[t]
\begin{center}\leavevmode
\epsfxsize=65mm \epsffile{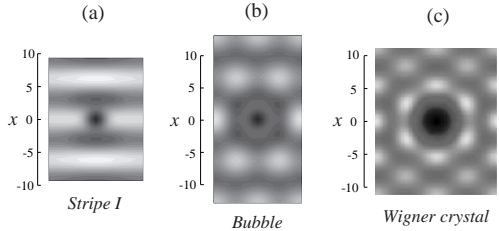}
\caption{Pair correlation function of (a) type I stripe state, 
(b) two-electron bubble state, and (c) Wigner crystal.}\label{CDW}
\end{center}\end{figure}

The ground state phase diagram in the third lowest Landau level 
is presented in Fig.~\ref{n2}\cite{Shib1}.
Between $\nu_N=1/2$ and $0.38$, the type I stripe state 
shown in Fig.~\ref{CDW} (a) is obtained. 
The amplitude of the stripes is much larger and the period is 
longer than those of  stripes in the second lowest Landau level.
Between $\nu_N =0.38$ and $0.24$, the two-electron bubble state 
shown in Fig.~\ref{CDW} (b) is found. 
Since the correlation function sharply changes at the
phase boundary, the transition between the stripe state and the bubble
state is expected to be first order. With further decreasing $\nu_N$, 
the bubble state becomes unstable and different ground state characterized
by Wigner crystal is realized\cite{Shib1}.

\section{Summary}
In the present paper we have studied the ground state of 
2D electrons in  Landau levels of $N=0,1,2$ 
by using the DMRG method.
From the analysis of the pair correlation functions and
low energy excitations, we have determined the phase diagram
consisting of various quantum liquids and CDW states. 
In particular we have found the stripe state in 
the lowest Landau level, which is realized only when the 
width of two-dimensional system is 
smaller than several times the magnetic length. 


\begin{thebibliography}{9}
\bibitem{period}
We note that the period of the stripes about $4\ell$
shown in Fig.~\ref{n0gr} (b) is different from
that predicted in the composite fermion theory 
of $10\ell$ \cite{CFBu} at $\nu=3/8$.
\end{thebibliography}
\end{document}